\documentclass[preprint,12pt]{elsarticle}
\usepackage{amssymb,amsmath,xcolor}

\begin{document}
\title{Non-Hermitian oscillator Hamiltonians and multiple Charlier polynomials}
\author{Hiroshi Miki\corref{cor1}}
\ead{miki@amp.i.kyoto-u.ac.jp}
\address{Department of Applied Mathematics and Physics, Graduate School of Informatics, Kyoto University, Sakyo-Ku, Kyoto 606 8501, Japan}

\author{Luc Vinet\corref{cor1}}
\ead{luc.vinet@umontreal.ca}
\address{Centre de recherches math\'{e}matiques, Universit\'{e} de Montr\'{e}al, P. O. Box 6128, Centre-ville Station, Montr\'{e}al (Qu\'{e}bec), H3C 3J7, Canada}
\author{Alexei Zhedanov\corref{cor1}}
\ead{zhedanov@fti.dn.ua}
\address{Donetsk Institute for Physics and Technology, Donetsk 83 114, Ukraine}

\begin{abstract}
A set of $r$ non-Hermitian oscillator Hamiltonians in $r$ dimensions is shown to be simultaneously diagonalizable.
Their spectra is real and the common eigenstates are expressed in terms of multiple Charlier polynomials. An algebraic interpretation of these
polynomials is thus achieved and the model is used to derive some of their properties.\\
\end{abstract}
\begin{keyword}
multiple Charlier polynomials \sep non-Hermitian Hamiltonians \sep
algebraic model.
\end{keyword}
\maketitle
\newpage
Non-Hermitian Hamiltonians with real spectra are currently being actively investigated with respect to their mathematical underpinnings
and their physical applications. (See the most recent special issues dedicated to this topic \cite{cite1,cite2,cite3}.)
We present here a set of $r$ non-Hermitian oscillator Hamiltonians in $r$ dimensions that all have real eigenvalues.
This system is also seen to provide an algebraic model for the multiple Charlier polynomials and is exploited to derive some properties
of these special functions. The approach is related to the interpretation that was given of the ordinary Charlier polynomials in \cite{cite4,cite5}
and of the $d$-orthogonal Charlier polynomials in \cite{cite6}.

The monic $r$-multiple Charlier polynomials \cite{cite7,cite8,cite9}  $C_{\vec{n}}^{\vec{\sigma }}(k)$
are indexed by a multi-index $\vec{n}=(n_1,n_2,\cdots ,n_r)\in \mathbb{N}^r$ with length $|\vec{n}|=n_1+\cdots +n_r$.
They are orthogonal with respect to $r$ Poisson measures with different positive parameters $\sigma_1,\cdots,\sigma_r$
(collectively denoted by $\vec{\sigma }$):
\begin{equation}
\sum_{k=0}^{\infty }C_{\vec{n}}^{\vec{\sigma }}(k)k^l\frac{\sigma_j^l}{k!}=0, \quad l=0,1,\cdots ,n_j-1
\end{equation}
for $1 \le j \le r$. (In the following we shall omit the suffix
$\vec{\sigma }$.) They have been shown \cite{cite7,cite10} to obey
the $r$ nearest-neighbor recurrence relations:
\begin{align}\label{eq1}
\begin{split}
kC_{\vec{n}}(k)&=C_{\vec{n}+\vec{e_1}}(k)+(\sigma_1+|\vec{n}|)C_{\vec{n}}(k)+\sum_{j=1}^rn_j\sigma_j C_{\vec{n}-\vec{e_j}}(k),\\
&\qquad \qquad \vdots \\
kC_{\vec{n}}(k)&=C_{\vec{n}+\vec{e_r}}(k)+(\sigma_r+|\vec{n}|)C_{\vec{n}}(k)+\sum_{j=1}^rn_j\sigma_j C_{\vec{n}-\vec{e_j}}(k),
\end{split}
\end{align}
where $\vec{e}_j=(0,\cdots, 0,1,0,\cdots ,0)$ is the $j$-th standard unit vector with $1$ on the $j$-th entry.
Subtracting the above relations pair-wise, one finds that the polynomials $C_{\vec{n}}(k)$ satisfy as a consequence
\begin{equation}\label{eq2}
C_{\vec{n}+\vec{e_i}}(k)-C_{\vec{n}+\vec{n_j}}(k)+(\sigma_i-\sigma_j) C_{\vec{n}}(k)=0
\end{equation}
for all $i,j \in \{ 1,\cdots ,r\}$.

Let us introduce the Heisenberg-Weyl algebra $W(r)$ associated to harmonic oscillators in $r$-dimensions.
$H$ is generated by the annihilation and creation operators $a_i$ and $a_i^+$ (resp.), $i=1,\cdots ,r$, that satisfy the commutation relations
\begin{equation}
[a_i,a_j]=[a_i^+,a_j^+]=0,\quad [a_i,a_j^+]=\delta_{ij},\quad i,j=1,\cdots,r.
\end{equation}
Denote by $\left| n_1,\cdots ,n_r\right>=\left| n_1\right>\cdots \left| n_r\right>$ the normalized simultaneous
eigenstates of the $r$ number operators $N_i=a_i^+a_i$:
\begin{align}
&\begin{aligned}
a_i^+a_i\left| n_1,\cdots,n_i,\cdots  ,n_r\right>&=n_i\left| n_1,\cdots,n_i,\cdots  ,n_r\right>, \\
&\qquad n_i\in \mathbb{N},\quad i=1,\cdots ,r,
\end{aligned}\\
&\left< m_1,\cdots ,m_r|n_1,\cdots ,n_r\right>=\delta_{m_1,n_1}\cdots \delta_{m_r, n_r}.
\end{align}
Remember that
\begin{equation}
[a_i^+a_i,a_j]=-\delta_{ij}a_j,\quad [a_i^+a_i,a_j^+]=\delta_{ij}a_j^+.
\end{equation}
The algebra $W(r)$ is represented in this number state basis in the standard way:
\begin{align} \label{aa_action}
\begin{split}
a_i\left| n_1,\cdots ,n_i,\cdots ,n_r\right>&=\sqrt{n_i} \left| n_1,\cdots ,n_i-1,\cdots ,n_r\right>,\\
a_i^+\left| n_1,\cdots ,n_i,\cdots ,n_r\right>&=\sqrt{n_i+1}\left| n_1,\cdots ,n_i+1,\cdots ,n_r\right>.
\end{split}
\end{align}

Consider now the set of $r$ Hamiltonians $H_i,i=1,\cdots,r$, defined as follows:
\begin{equation}
H_i=\sum_{j=1}^ra_j^+a_j+\sum_{j=1}^r\sigma_ja_j^+ +a_i+\sigma_i,
\quad i=1,\cdots ,r.
\end{equation}
It is straightforward to see that the multiple Charlier polynomials simultaneously diagonalize the $r$ non-Hermitian oscillator Hamiltonians.
To that end, form the states
\begin{equation}\label{eq9}
\left.\left| k\right>\right> = N_k^{(r)}\sum_{\vec{n}=0}^{\infty }\frac{C_{\vec{n}}(k)}{\sqrt{n_1!\cdots n_r!}}\left| n_1,\cdots ,n_r\right>,\quad k\in \mathbb{N}.
\end{equation}
Let us act on $\left.\left| k\right>\right>$ with $H_i$:
\begin{align}
\begin{split}
H_i\left.\left| k\right>\right> =&N_k^{(r)}\sum_{\vec{n}=0}^{\infty }\frac{C_{\vec{n}}(k)}{\sqrt{n_1!\cdots n_r!}} \\
&\cdot \Biggl\{ (|\vec{n}|+\sigma_i) \left| n_1,\cdots ,n_i,\cdots ,n_r\right>  \\
&+\sqrt{n_i}\left| n_1,\cdots ,n_i-1,\cdots ,n_r\right>  \\
&\left. +\sum_{j=1}^r\sigma_j\sqrt{n_j+1}\left| n_1,\cdots ,n_i+1,\cdots ,n_r\right>
\right\} \\
=&N_k^{(r)}\sum_{\vec{n}=0}^{\infty }\frac{1}{\sqrt{n_1!\cdots n_r!}}\\
&\cdot \left\{
C_{\vec{n}+\vec{e_i}}(k)+(\sigma_i+|\vec{n}|)C_{\vec{n}}(k)+\sum_{j=1}^rn_j\sigma_j
C_{\vec{n}-\vec{e_j}}(k)\right\},
\end{split}
\end{align}
Invoking the recurrence relations \eqref{eq1}, we thus have indeed
\begin{equation}
H_i\left.\left| k\right>\right> =kH_i\left.\left| k\right>\right>
\end{equation}
for all $i=1,\cdots ,r$. So although non-Hermitian, the operators
$H_i$ have a real spectrum given by the non-negative integers, the
states $\left.\left| k\right>\right>$ are uniquely defined (up to
a constant factor) as the joint eigenstates of the operators
$H_i,i=1,\cdots ,r$ with eigenvalues equal to $k$. Interestingly
the Hamiltonians $H_i$ do not commute pairwise. Indeed it is
readily found that
\begin{equation}
[H_i,H_j]=a_i-a_j+(\sigma_i-\sigma_j)
\end{equation}
for all $i,j=1.\cdots ,r$. Remark however that
\begin{align}
\begin{split}\label{eq10}
(a_i-a_j)\left.\left| k\right>\right>
=&N_k^{(r)}\sum_{\vec{n}=0}^{\infty }\frac{C_{\vec{n}}(k)}{\sqrt{n_1!\cdots n_r!}} \\
&\cdot \biggl\{ \sqrt{n_i}\left| n_1,\cdots ,n_i-1,\cdots ,n_r\right> \\
&-\sqrt{n_j}\left| n_1,\cdots ,n_i-1,\cdots ,n_r\right> \biggr\} \\
=&N_k^{(r)}\sum_{\vec{n}=0}^{\infty }\frac{1}{\sqrt{n_1!\cdots n_r!}} \\
&\cdot \left(C_{\vec{n}+\vec{e}_i}(k)-C_{\vec{n}+\vec{e}_j}(k)\right)\left| n_1,\cdots ,n_r\right>
\end{split}
\end{align}
Hence, by property \eqref{eq2},
\begin{equation}
[H_i,H_j]\left.\left| k\right>\right>=0.
\end{equation}
The operators $H_i$, thus on commute ``on shell", thereby reconciling the fact that they do not commute and yet have common eigenvectors.
As such the set of Hamiltonians $H_i$ form
a ``weakly" integrable system. It is also useful to observe that the operators $H_i$ can be obtained from the standard harmonic oscillator
Hamiltonian in $r$ dimensions
\begin{equation}
H_0=\sum_{j=1}^ra_j^+a_j
\end{equation}
by similarity transformations. Using the Baker-Campbell-Hausdorff
formula
\begin{equation}
e^{A}Ye^{-B}=A+[A,B]+\frac{1}{2!}[A,[A,B]]+\frac{1}{3!}[A,[A[A,B]]]+\cdots ,
\end{equation}
it is easily shown that
\begin{equation}\label{eq3}
H_i=S_iH_0S_i^{-1}
\end{equation}
with
\begin{equation}\label{eq4}
S_i=e^{a_i}\prod_{j=1}^re^{-\sigma_ja_j^+}, \quad i=1,\cdots ,r.
\end{equation}
This explains obviously why the spectrum of the $H_i$ are real.

We shall now show how this framework can be used to derive properties of the multiple Charlier polynomials.
We shall focus on the step relations they obey and their explicit expression.
Suppose we have operators $X$ such that
\begin{equation}\label{eq5}
[H_i,X] =-X
\end{equation}
for all $i=1,\cdots ,r$. Then,
\begin{equation}
H_i\left(X\left.\left| k\right>\right>\right)=(k-1)X\left.\left| k\right>\right>,
\end{equation}
that is, $X\left.\left| k\right>\right>$ is a simultaneous eigenvectors of the $H_i$ with eigenvalue $k-1$
and is hence proportional to $\left.\left| k-1\right>\right>$. Similarly, if there is an operator Y such that
\begin{equation}\label{eq8}
[H_i,Y]=Y
\end{equation}
for all $i=1,\cdots ,r$, the vector $Y\left.\left| k\right>\right>$ will be proportional to $\left.\left| k+1\right>\right>$.
It is immediate to find these operators $X$ and $Y$ from the known lowering and raising operators of $H_0$,
by exploiting the fact that the $H_i$'s  are related to $H_0$ by the similarity transformations \eqref{eq3}-\eqref{eq4}.
Let $X_0=\sum_{i=1}^r\alpha_ia_i$, where $\alpha_i$ are some constants. It is clear that $[H_0,X_0]=-X_0$.
To obtain $X$s that will obey \eqref{eq5}, one has to determine the constants $\alpha_i$ so that
\begin{equation} \label{eq6}
S_1X_0S_1^{-1}=S_2X_0S_2^{-1}=\cdots =S_rX_0S_r^{-1}.
\end{equation}
In this case, it is found that \eqref{eq6} imposes no constraints on the $\alpha_i$s and hence the $r$ operators
\begin{equation}\label{eq7}
X_j=S_ia_jS_i^{-1}=a_j+\sigma_j
\end{equation}
enjoy the property \eqref{eq5}. They will be found to imply $r$ step relations for the multiple Charlier polynomials.
Similarly, using $Y_0=\sum_{i=1}^r \beta_i a_i^+$ as the generic raising operator of $H_0$,
the condition analogous to \eqref{eq7} is found to require that all the coefficients $\beta_i$ be equal.
Up to a trivial constant factor there is thus only one raising operator $Y$ verifying \eqref{eq8} and it is given by
\begin{equation}\label{eq13}
Y=S_i\left(\sum_{j=1}^ra_j^+\right) S_i=\left(\sum_{j=1}^ra_j^+\right)+1.
\end{equation}
In order to obtain step relations for the multiple Charlier
polynomials from the action of $X_j, j=1,\cdots ,r$ and $Y$ on
both sides of \eqref{eq9}, we need to know the precise action of
these operators on $\left.\left| k\right>\right>$. The key is to
relate $\left.\left| k\right>\right>$ to an eigenstate of $H_0$.

Let $\left| k\right>^*$ be a state such that
\begin{equation}\label{eq12}
H_0\left| k\right>^*=k\left| k\right>^*.
\end{equation}
Take $S_1\left| k\right>^*$. Obviously,
\begin{equation}
H_1S_1\left| k\right>^*=S_1H_0\left| k\right>^*=kS_1\left| k\right>^*.
\end{equation}
That is, $S_1\left| k\right>^*$ is an eigenstate of $H_1$ with eigenvalue $k$. If now,
\begin{equation}\label{eq15}
S_i\left| k\right>^*=\gamma _iS_1\left| k\right>^*, \quad
i=2,\cdots ,r
\end{equation}
for all $i\ne 1$, with $\gamma _i$ some constants,
because of \eqref{eq3}, $S_1\left| k\right>^*$ will necessarily be a common eigenstate of all $H_i,i=1,\cdots,r$, with eigenvalue $k$.
Since this last property defines $\left.\left| k\right>\right>$ (up to a constant factor), we could then posit
\begin{equation}\label{eq11}
\left.\left| k\right>\right>=S_1\left| k\right>^*.
\end{equation}
Let us then identify the state $\left.\left| k\right>\right>$ for
which \eqref{eq11} will hold. It is determined by the conditions
\eqref{eq12} and \eqref{eq15}. The most general $\left|
k\right>^*$ satisfying \eqref{eq12} will be of the form
\begin{equation}\label{eq14}
\left| k\right>^*=M_k^{(r)} \sum_{l_2,\cdots,l_r} d_{l_2,\cdots ,l_r} \left| k-|\vec{l}|,l_2,\cdots ,l_r\right>,
\end{equation}
where $d_{l_2,\cdots ,l_r}$ are some coefficients to be specified from \eqref{eq13} and $M_k^{(r)}$ is the normalization factor.
The sum in \eqref{eq14} is performed over all $l_i\in \{ 0,\cdots,k\}$ such
that $|\vec{l}|=(l_2+\cdots +l_r)\le k$.

Using the property $e^{A}e^B=e^{[A,B]}e^Be^A$ which is valid when the operators $A$ and $B$ commute with their commutator,
one readily finds from \eqref{eq4} that
\begin{equation}
S_i=e^{\sigma_1-\sigma_i}S_ie^{-a_1}e^{a_i},\quad i=2,\cdots ,r.
\end{equation}
If we choose the coefficients $\{ d_{l_2,\cdots ,l_r}\}$ so that
\begin{equation}\label{eq16}
e^{-a_1}e^{-a_i}\left| k\right>^*=\left| k\right>^*
\end{equation}
for all $i=2,\cdots,r$, condition \eqref{eq15} will be satisfied with $\gamma _i=e^{\sigma_1-\sigma_i}$. Now \eqref{eq16}
is tantamount to demanding that
\begin{equation}\label{eq17}
(a_1-a_i)\left| k\right>^*=0,\quad i=2,\cdots ,r.
\end{equation}
The $r-1$ constraints \eqref{eq17} are straightforwardly found to imply the following $r-1$ recurrence relations
for the coefficients $\{ d_{l_2,\cdots ,l_r}\}$:
\begin{equation}
\sqrt{l_i}d_{l_2,\cdots l_i+1,\cdots ,l_r}=\sqrt{k-|\vec{l}|}d_{l_2,\cdots l_i,\cdots ,l_r}
\end{equation}
which are solved by
\begin{equation}
d_{l_2,\cdots ,l_r} = \sqrt{\frac{k!}{(k-|\vec{l}|)!l_2!\cdots l_r!}}d_{0,\cdots ,0}.
\end{equation}
Hence
\begin{equation}\label{eq31}
\left| k\right> ^*=\tilde{M}_k^{(r)} \sum_{l_2,\cdots,l_r} \sqrt{\frac{k!}{(k-|\vec{l}|)!l_2!\cdots l_r!}}\left| k-|\vec{l}|,l_2,\cdots ,l_r\right>,
\end{equation}
where $\tilde{M}_k^{(r)}=M_k^{(r)}d_{0,\cdots,0}$. This normalization factor is readily determined from the condition that $\left| k\right>^*$
be normalized: ${}^*\left< k|k\right>^*=1$. This yields
\begin{equation}\label{eq32}
\tilde{M_k}^{(r)}=\frac{1}{r^{\frac{k}{2}}}
\end{equation}
with the help of the multinomial formula. The action of the lowering and raising operators of $\left| k\right> ^*$ can now be directly computed.
One finds that
\begin{equation}\label{eq18}
a_i\left| k\right> ^*=\sqrt{\frac{k}{r}}\left| k-1\right> ^*
\end{equation}
and also that
\begin{equation}\label{eq21}
(a_1^++a_2^++\cdots +a_r^+)\left| k\right> ^*=\sqrt{r(k+1)}\left| k+1\right> ^*.
\end{equation}
Although, we have privileged the use of $S_1$ in the preceding considerations,
this choice must not matter and indeed the vector $\left| k\right>^*$ that has been found is seen to symmetrically satisfy the constraints
\begin{equation}\label{eq26}
(a_i-a_j)\left| k\right> ^*=0
\end{equation}
for all pair $i,j=1,\cdots ,r$.

We are now ready to complete the derivations of the step relations. Using \eqref{eq7}, \eqref{eq11} and \eqref{eq18}, we can now write that
\begin{equation}
(a_j+\sigma_j) \left.\left| k\right>\right> =\left.\left| k-1\right>\right>.
\end{equation}
Using \eqref{eq9}, this yields
\begin{equation}\label{eq19}
\sqrt{\frac{k}{r}}N_{k-1}^{(r)}C_{\vec{n}}(k-1)=N_k^{(r)}(C_{\vec{n}+\vec{e_j}}(k)+\sigma_j C_{\vec{n}}(k)).
\end{equation}
To proceed further we need to find $N_k^{(r)}$. In order to do that, set $\vec{n}=0$ in \eqref{eq19}
and use the recurrence relations \eqref{eq1} to observe that
\begin{equation}
C_{\vec{e_j}(k)}=k-\sigma_j.
\end{equation}
We thus have
\begin{equation}\label{eq20}
\frac{N_{k-1}^{(r)}}{N_{k}^{(r)}}=\sqrt{rk}
\end{equation}
which implies
\begin{equation}\label{eq22}
N_k^{(r)}=\frac{1}{\sqrt{k!}r^{\frac{k}{2}}}N_0
\end{equation}
with
\begin{equation}
N_0=\left< 0,\cdots ,0|S_1|0,\cdots ,0\right>=e^{-\sigma_1}.
\end{equation}
 Returning to \eqref{eq19}, we finally have using \eqref{eq20}, the following set of $r$ ``backward'' step relations:
\begin{equation}\label{eq23}
kC_{\vec{n}}(k-1)=C_{\vec{n}+\vec{e_j}}(k)+\sigma_jC_{\vec{n}}(k),\quad
j=1,\cdots ,r.
\end{equation}
Similarly, for the raising operator, we see from \eqref{eq13}, \eqref{eq11} and \eqref{eq21} that we have
\begin{equation}\label{eq24}
(a_1^++\cdots +a_r^++1)\left.\left| k\right>\right>=\sqrt{r(k+1)}\left.\left| k+1\right>\right>.
\end{equation}
Using again the coherent sum \eqref{eq9}, with $N_k^{(r)}$ now given by \eqref{eq22},
steps analogous to those that led to \eqref{eq23} bring one to find the ``forward" relation that \eqref{eq24} entail:
\begin{equation}\label{eq25}
C_{\vec{n}}(k+1)=C_{\vec{n}}(k)+\sum_{j=1}^nn_jC_{\vec{n}-\vec{e_j}}(k).
\end{equation}
The relation \eqref{eq23} and \eqref{eq25} can obviously be combined to obtain a difference equation for the multiple polynomials $C_{\vec{n}}^{k}$.

The Hamiltonian $H_0$ is known to admit a $U(r)$ invariance
algebra, generated by the bilinears $a_i^+a_j,i,j=1,\cdots ,r$
that all commute with $H_0$. One might be curious to know what is
their inference on the discrete integrable system on the lattice
$\mathbb{N}^r$ for which the multiple Charlier polynomials provide
solutions. As for the joint ladder operators $X$ and $Y$, the
common symmetries $R$ of the $H_i$ ($[H_i,R]=0,i=1,\cdots ,r$.)
are obtained from the symmetries $R_0$ of $H_0$ that have all
their conjugates $S_iR_0S_i^{-1},i=1,\cdots ,r$ differ at most by
a constant (that does not matter). Resolving those conditions, it
is found that the common symmetries must be of the form
\begin{align}
\begin{split}
R_{ij}&=-R_{ji}
=S_1[(a_1^++\cdots +a_r^+)(a_i-a_j)]S_1^{-1}\\
&=(a_1^++\cdots +a_r^++1)(a_i+\sigma_i-a_j-\sigma_j)
\end{split}
\end{align}
From \eqref{eq26}, it is seen that they all annihilate $\left.\left| k\right>\right>$ owing to the factor $(a_i-a_j)$
in their expression in the first line. Note that $S_1(a_1-a_j)S_1^{-1}=(a_i+\sigma_i -a_j-\sigma_j)$.

Observe also that these joint constants of motion are in
involution:
\begin{equation} \label{Rcom}
[R_{ij},R_{kl}]=0.
\end{equation}

By making explicit, the implication of
\begin{equation}
R_{ij}\left.\left| k\right>\right>=0
\end{equation}
when $R_{ij}$ acts on the r.h.s. of \eqref{eq9}, one obtains the relations
\begin{align}
\begin{split}
&\sum_{s\ne i} n_iC_{\vec{n}+\vec{e_s}-\vec{e}_i}(k)-\sum_{s\ne j} n_jC_{\vec{n}+\vec{e_s}-\vec{e}_j}(k)\\
&+(n_i-n_j+\sigma_i-\sigma_j)C_{\vec{n}}(k)+C_{\vec{n}+\vec{e_i}}(k)-C_{\vec{n}+\vec{e_j}}(k)\\
&+(\sigma_i -\sigma_j) \sum_{s=1}^rn_s C_{\vec{n}-\vec{e}_s}(k)=0
\end{split}
\end{align}
which are in fact consequences of \eqref{eq2} when combined with the recurrence relations \eqref{eq1}.

Our final considerations will bear on the explicit expression of
the multiple Charlier polynomials and on the generating function.
Given \eqref{eq11}, that is $\left.\left| k\right>\right> =
S_1\left| k\right>^*$, and the fact that $S_1$ and $\left|
k\right>^*$ are known explicitly, it is quite clear that an
expression for the multiple Charlier polynomials should follow
from working out the action of $S_1$ on $\left| k\right>^*$. Let
us then proceed.
\begin{align}
&S_1=e^{-\sigma_1}e^{-\sigma_1a_1^+}e^{a_1}\cdot e^{-\sigma_2 a_2^+}\cdots e^{-\sigma _r a_r^+}, \\
&\begin{aligned}\label{eq28}
S_1&\left| k-|\vec{l}|,l_2,\cdots ,l_r\right>\\
&=e^{-\sigma_1}e^{-\sigma_1a_1^+}e^{a_1}\left| k-|\vec{l}| \right> \cdot \prod_{j=2}^r e^{-\sigma_j a_j^+}\left| l_j\right>,
\end{aligned}\\
&\begin{aligned}\label{eq29}
e^{-\sigma_j a_j^+}\left|l_j\right> &= \sum_{n_j=0}^{\infty } \left| n _j\right> \left< n_j \right| e^{-\sigma_ja_j^+}\left| l_j\right>\\
&=\sum_{n_j=0}^{\infty }\frac{(-\sigma )^{n_j-l_j}}{(n_j-l_j)!}\sqrt{\frac{n_j!}{l_j!}}\left| n_j \right>,
\end{aligned}\\
&e^{-\sigma_1 a_1^+}e^{a_1}\left| k-|\vec{l}|\right> =\sum_{n_1=0}^{\infty } \left| n _1\right> \left< n_1\right| e^{-\sigma_1 a_1^+}e^{a_1}\left| k-|\vec{l}|\right>. \label{eq27}
\end{align}
The matrix elements occurring in \eqref{eq27} can be obtained following techniques used in \cite{cite4} and \cite{cite5}. If one sets
\begin{equation}
\psi_{n,k}=\left< k \right| e^{-\sigma_1a_1^+} e^{a_1} \left|
n\right>,
\end{equation}
from
\begin{equation}
\left< k \right| a_1^+a_1e^{-\sigma_1a_1^+}e^{a_1} \left| n\right>
=\left< k \right|
e^{-\sigma_1a_1}e^{a_1}(a_1^+a_1-a_1-\sigma_1a_1^++\sigma_1)\left|
n \right>,
\end{equation}
one readily finds that $\psi_{n,k}$ obey
the recurrence relation
\begin{equation}
k\psi_{n,k}=(n+\sigma_1)\psi_{n,k}-\sqrt{n}\psi_{n-1,k}-\sigma_1\sqrt{n+1}\psi_{n+1,k}.
\end{equation}
This implies that
\begin{equation}
\psi_{n,k}=\frac{1}{\sqrt{n!}}(-\sigma_1)^np_n(k)\psi_{0,k},
\end{equation}
where $p_n(k)$ are polynomials of order $n$ in $k$ which are identified with the monic Charlier polynomials.
The ``initial'' value $\psi_{0,k}$ is directly computed to be
\begin{equation}
\psi_{0,k}=\frac{(-\sigma_1)^k}{\sqrt{k!}}.
\end{equation}
This therefore implies that
\begin{equation}\label{eq30}
e^{-\sigma_1a_1^+} e^{a_1}\left| k-|\vec{l}| \right>
=\sum_{n_1=0}^{\infty
}\frac{(-\sigma_1)^{n_1+|\vec{l}|-k}}{\sqrt{n_1!(k-|\vec{l}|)!}}p_{k-|\vec{l}|}(n_1)\left|
n_1\right>.
\end{equation}
One then uses \eqref{eq28}, \eqref{eq29} and \eqref{eq30} to write down the action of $S_1$ on $\left| k\right>^*$
as given by \eqref{eq31} and \eqref{eq32}. First, one calls upon the
following expression for the standard Charlier polynomials \cite{cite11}
\begin{align}
\begin{split}
p_n(k)&=(-\sigma_1)^n ~_2F_0
\left(
\begin{matrix}
-n,-k\\
-
\end{matrix}\left. \right|-\frac{1}{\sigma_1}
\right) \\
&=(-\sigma_1)^n \sum_{s=0}^{\infty } \frac{(-n)_s(-k)_s}{s!}\left( -\frac{1}{\sigma_1}\right)^s,
\end{split}
\end{align}
where we are using the Pochhammer symbol defined by
\begin{equation}
(a)_r=a(a+1)\cdots (a+r-1).
\end{equation}
One then identifies the coefficients of $\left| n_1,\cdots ,n_r\right>$ on both sides of $\left.\left| k \right>\right>=S_1\left| k\right>^*$
to find that
\begin{align}
\begin{split}
 C_{\vec{n}}(k)=\sum_{s=0}^{n_1}\sum_{\{ l_i\}} &\frac{k!(|\vec{l}|-k)_r}{(k-|\vec{l}|)!}\cdot (-n_1)_r\cdot \frac{n_2!\cdots n_r!}{(n_2-l_2)!\cdots (n_r-l_r)!}\\
&\cdot \frac{(-\sigma_1)^{n_1-s}}{s!} \cdot \frac{(-\sigma_2)^{l_2-s}}{l_2!}\cdots \frac{(-\sigma_r)^{l_r-s}}{l_r!}.
\end{split}
\end{align}
After some simplification, one finally arrives at the following explicit expression:
\begin{align}
\begin{split}
C_{\vec{n}}(k)=\sum_{l_1=0}^{n_1}\cdots \sum_{l_r=0}^{n_r}&(-n_1)_{l_1}(-n_2)_{l_2}\cdots (-n_r)_{l_r}(-k)_{l_1+\cdots +l_r} \\
&\cdot \frac{(-\sigma_1)^{n_1-l_1}}{l_1!} \cdot \frac{(-\sigma_2)^{n_2-l_2}}{l_2!}\cdots \frac{(-\sigma_r)^{n_r-l_r}}{l_r!}.
\end{split}
\end{align}
This formula has been given in \cite{cite7} for the case $r=2$.

Last, a generating function can be obtained from \eqref{eq11}, by
introducing the Bargmann representation where the states $| n_1,
n_2, \dots, n_t \rangle$ and the oscillator operators $a_i,
a_i^{+}, \; i=1,2,\dots,r$ are realized by
\begin{equation} \label{Barg}
| n_1, n_2, \dots, n_r \rangle  = \frac{1}{\sqrt{n_1! n_2! \dots
n_r!}} z_1^{n_1} z_2^{n_2} \dots z_r^{n_r}, \quad
a_i=\partial_{z_i}, \: a^{+}_i =z_i
\end{equation}
Relations \eqref{aa_action} are obviously enforced in this model.

Recall from \eqref{eq15} and \eqref{eq11} that
\begin{equation} \label{ess}
e^{\sigma_i-\sigma_1} S_i |k\rangle^* =|k\rangle \rangle
\end{equation}
for any $i=1,2,\dots,r$. In the Bargmann representation
\begin{equation}
S_i = e^{-\sigma_1} e^{-(\sigma_1 z_1 + \dots + \sigma_r z_r)}
e^{\partial_{z_i}}
\end{equation}
and
\begin{equation} \label{k_sum}
|k \rangle^*= \sqrt{\frac{k!}{r^k}} \sum_{l_1+l_2+\dots + l_r=k}
\frac{z^{l_1}}{l_1!} \dots  \frac{z^{l_r}}{l_r!}
\end{equation}
Since
$$
e^{\partial_{z_i}} f(z_1, z_2, \dots, z_i, \dots, z_r) = f(z_1,
z_2, \dots, z_i+1, \dots, z_r),
$$
we have, combining \eqref{ess} and \eqref{k_sum}
\begin{align}
\begin{split}
\frac{e^{-\sigma_1}\sqrt{k!}}{r^{k/2}} e^{-(\sigma_1 z_1 + \dots +
\sigma_r z_r)} \sum_{l_1, \dots l_r}\frac{z_1^{l_1} \dots
(z_i+1)^{l_i} z_r^{l_r}
}{l_1! \dots l_r!} = \\
\frac{e^{-\sigma_1}}{\sqrt{k! r^k}} \sum_{n_1, \dots n_r}
\frac{C_{\vec n}(k)}{n_1! \dots n_r!} z_1^{n_1} \dots z_r^{n_r}.
\end{split}
\end{align}
This readily gives, the following generating function with the
help of the multimonomial formula
\begin{align} \label{gen}
\begin{split}
e^{-(\sigma_1 z_1 + \dots + \sigma_r z_r)} (z_1+z_2 + \dots +z_r
+1)^k = \\
\sum_{\vec n} C_{\vec n}(k) \frac{z_1^{n_1}}{n_1!} \dots
\frac{z_r^{n_r}}{n_r!}
\end{split}
\end{align}
Notice that, as should be, any operator $S_i, \; i=1, \dots, r$
can be used to derive the identity \eqref{gen}.

In summary, let us recall some of our findings. We have presented
a remarkable set of non-Hermitian harmonic oscillator Hamiltonians
in $r$-dimensions with real spectra. Their common eigenfunctions
have been seen to be given in terms of multiple Charlier
polynomials. Had we passed to a coordinate representation, we
could have obtained relations involving products of Hermite
polynomials and their translates \cite{cite4,cite5}. This physical
settings has provided an algebraic model for the multiple Charlier
polynomials that has been used to offer alternate demonstrations
of some of their structural relations. It would be of interest in
our opinion to pursue the algebraic interpretation of multiple
polynomials. The multiple Meixner polynomials in particular,
should lend themselves also to an oscillator modelization. We plan
to return to this in a future publication.

\section*{Acknowledgements}
One of us (H.M.) would like to thank the CRM for its hospitality
while this work was carried out. His work is supported by a
Grant-in-Aid for Japan Society for the Promotion of Science (JSPS)
Fellows. The research of (L.V.) is supported in part through funds
provided by the National Sciences and Engineering Research Council
(NSERC) of Canada.

\end{document}